\documentclass{spie}
\usepackage{amsmath,amssymb}
\usepackage{graphicx}
\newcommand{\re}[1]{(\ref{#1})}
\newcommand{\lsim}{\stackrel{\scriptstyle <}{\phantom{}_{\sim}}}

\begin{document}

\title{Observable effects caused by vacuum pair creation in the field
of high-power optical lasers}

\author{David B. Blaschke\supit{a,b,c}, Andrey V. Filatov\supit{d},
 Irina A. Egorova\supit{b,d}, \\ Alexander V. Prozorkevich\supit{d}
and Stanislav A. Smolyansky\supit{d} \skiplinehalf \supit{a}
Institute of Theoretical Physics, University of
Wroc{\l}aw, 50-204 Wroc{\l}aw, Poland \\
\supit{b} Bogoliubov Laboratory for Theoretical Physics, JINR Dubna, RU-141980 Dubna, Russia\\
\supit{c} Institute of Physics, University of Rostock, D-18051 Rostock, Germany\\
\supit{d} Saratov State University, RU-410026, Saratov, Russia}

\maketitle

\begin{abstract}
We consider the possibility of an experimental proof of vacuum
$e^+e^-$ pair creation in the focus of two counter-propagating
optical laser beams with an intensity of the order of
10$^{20}$-10$^{22}$ W/cm$^2$. Our approach is based on the
collisionless kinetic equation for the distribution function of
the $e^+e^-$ pairs with the source term for particle production.
As a possible experimental signal of vacuum pair production we
consider the refraction of a high-frequency probe laser beam by
the produced $e^+e^-$ plasma to be observed by an interference filter.
The generation of higher harmonics of the laser frequency in the
self-consistent electric field is also investigated.
\end{abstract}


\section{Introduction}

The great progress in laser technology expected in near future
\cite{Tajama,BulanovSV03} raises hopes for a sought-after experimental
proof of some non-linear effects in quantum electrodynamics,
in particular, the dynamical Schwinger effect of vacuum pair creation
predicted a long time ago \cite{Sauter,HE,Schwinger}.

As is known \cite{Schwinger} no pairs can be created when both
invariants of the electromagnetic field vanish, $\mathbf{E}^2 -
\mathbf{B}^2=0$,  $\mathbf{E}\mathbf{B}=0$.
The field produced by focussed laser beams is very close to such
a configuration \cite{Troup} and therefore the pair creation is
expected to be essentially suppressed.
On the other hand, it should be possible to avoid the lightlike
field configuration with a spatially uniform field created in an
antinode of the standing wave produced by the superposition of
two coherent, counter-propagating laser beams \cite{Marinov}.
Since the Schwinger effect is non-perturbative and it requires an
exact solution of the dynamical equations it is customary to
approximate the complicated structure of a real laser field by a
spatially uniform time-dependent electric field.
According to early estimates
\cite{Richards,Bunkin,Brezin,Popov02,Marinov,Troup,BulanovSS05}
the effect of vacuum creation should not be observable with
presently available laser parameters.
Essentially, these estimations concerned the pairs which remain after
the laser pulse when the electric field disappears.
Technically, time interval considered has to be an integer multiple
of the field period.
Another possibility was investigated in the works
\cite{2004b,2006a,2006b,2006c,2006d}, where the dynamical pair density
during the action of laser pulse changes periodically with twice
the field frequency and its mean value exceeds the residual
density by several order of magnitude.
In that case, plenty of quasiparticle $e^+e^-$ pairs are created
even in weak time-dependent fields $\omega^2 \ll E \ll m^2/e$.
Though these pairs disappear together with driving field, they can
interact in standard ways (scattering, annihilation etc.) during the
period of field presence some traces of these processes should be revealed.

The nonlinear effects due to the interaction of intense laser beams with
the vacuum can manifest themselves in various physical phenomena, e.g.,
by nonlinear Thomson scattering \cite{Esarey}, damping of
electromagnetic waves in a plasma due to $e^+e^-$ pair production
\cite{BulanovSS05}, two-photon annihilation \cite{Landau4}, the
non-linear Breit-Wheeler process \cite{Ivanov}, vacuum
birefringence \cite{Heinzl} and dichroism \cite{Heyl}.
Some other interesting possibilities are considered recently in Refs.
\cite{Piazza,Kharzeev06}.
The estimation of the two-photon annihilation rate \cite{2006a} shows that
the field of an optical laser with intensity $10^{20}$~W/cm$^2$ can
produce  5-10 annihilation events per laser pulse with irradiation
of $\sim$ 1 MeV $\gamma$-quanta. The corresponding $\gamma$-quanta
can be registered outside the focus of the counter-propagating
laser beams.

In present work, we consider whether it is possible to observe
experimentally a refraction index of the transient $e^+e^-$ plasma created
in the laser focus by means of some interference scheme with a
high-frequency probe laser.
In the collision of two counter-propagating laser beams form thin spatial
regions with quasi-homogeneous electric field where conditions for
$e^+e^-$ pair creation are fulfilled.
One of the split beams of the probe laser passes through the plasma region
and obtains a phase shift relative to the second beam that can be measured
by an interference filter.
The density of created pairs oscillates with the double laser frequency.
Therefore, a stable interference pattern can be obtained provided the probe
laser pulse is short enough.
A simple estimate shows that the intensity of  state-of-the-art optical lasers
is not sufficient for such an experiment but the critical intensity can be
reached in near future.
The  $e^+e^-$ pairs moving in the laser field generate a secondary
electric field which contains higher harmonics of the laser frequency, the
most intense of which is third one.

\section{Model}

\begin{figure}[t]
\centering
\includegraphics[width=0.47\textwidth,keepaspectratio=true]{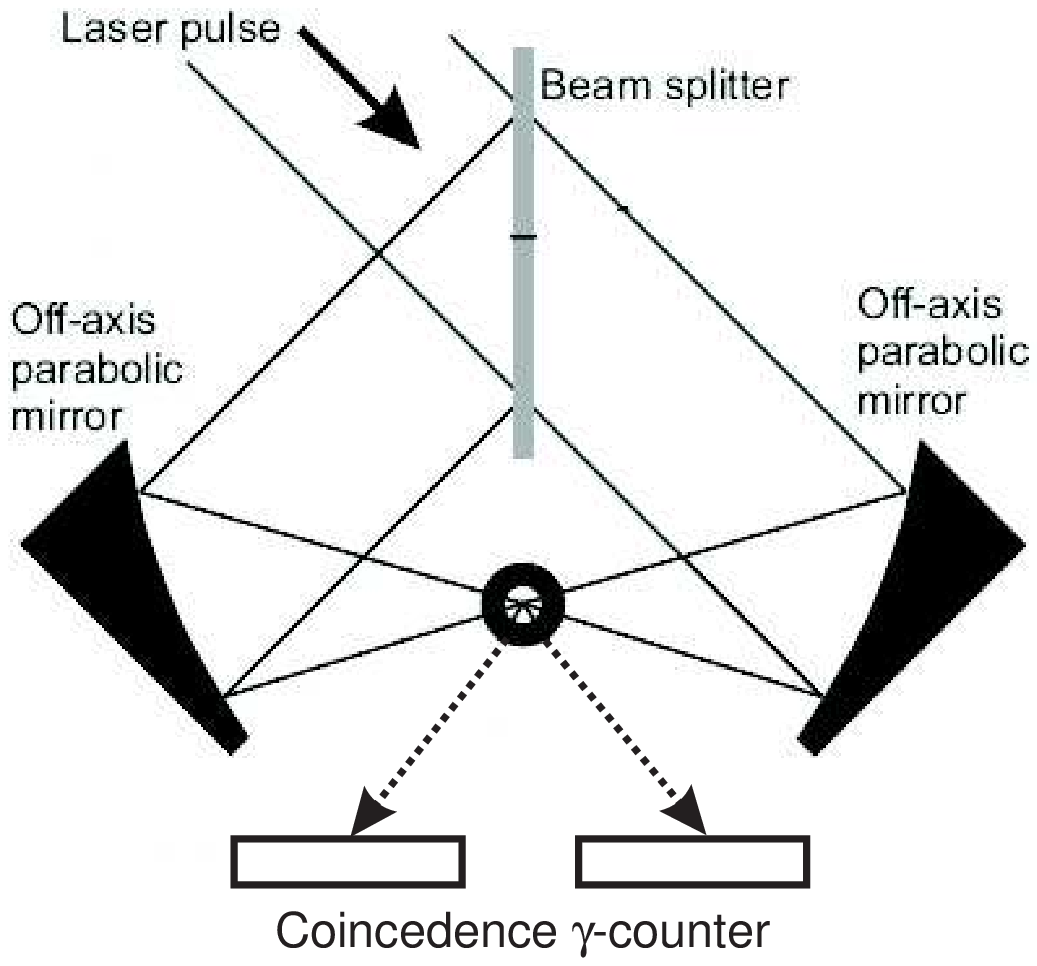}
\hfill
\includegraphics[width=0.47\textwidth,keepaspectratio=true]{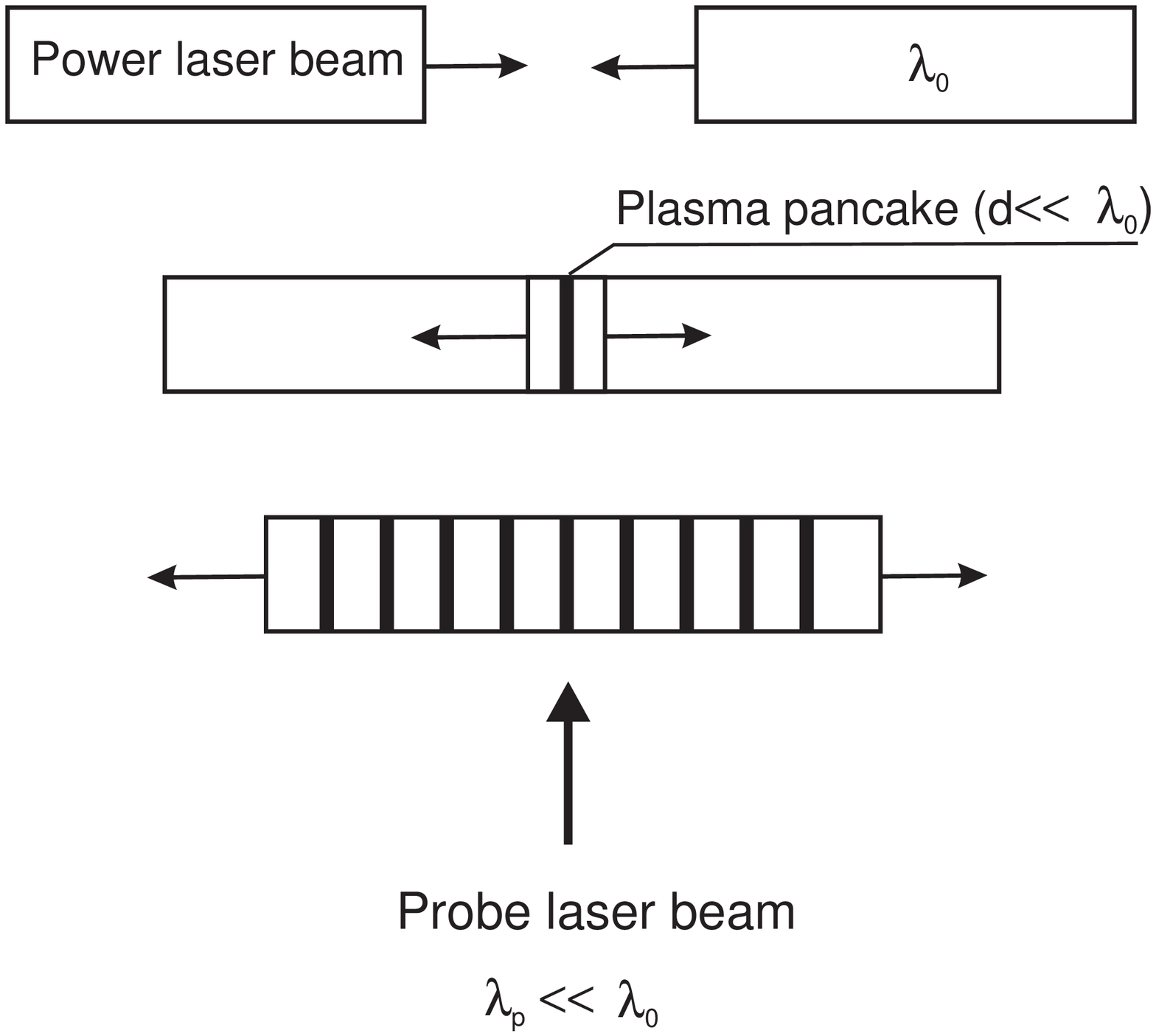}
\parbox[t]{0.47\textwidth}{
\caption{ Schematic representation of the experimental setup for two
colliding laser pulses \cite{Liesfeld05}. The main pulse of power laser is
divided into two pulses of equal energy by a beam splitter, both
pulses are focused in one point by a $45^{\circ}$ parabolic mirror.}
\label{fig1} }\hfill
\parbox[t]{0.47\textwidth}{
\caption{Formation of thin plasma regions ("pancakes") in the
collision of two counter-propagating laser beams with wavelength
$\lambda_0$. The beam of the probe laser has a short wavelength
$\lambda_p \ll \lambda_0$ and propagates perpendicular to the
direction of the high-power laser beams.}\label{fig2} }
\end{figure}

We suppose that two counter-propagating laser beams (Fig.\ref{fig1})
form a standing-wave field
\begin{align}
    \mathbf{E}= \{0, 0, 2E_m \phi(t)\sin{(\omega_0 t)}\cos{(kx)}\},
    \nonumber \\
    \mathbf{B}= \{0, 2E_m \phi(t) \cos{(\omega_0 t)}\sin{(kx)}, 0\},
\end{align}
where $\phi(t)$ is some envelope function. The small spatial regions
in the vicinity of the anti-nodes of the electric field (Fig. \ref{fig2})
can be considered as areas with quasi-homogeneous electric field
\begin{eqnarray}
  \mathbf{E}_{ex}(t) = (0,0,E_{ex}(t)),\qquad \qquad
  E_{ex}(t)= E_0 \phi(t) \sin{\omega_0 t}, \label{s1}
  \end{eqnarray}
which are capable of pair creation.
The Dirac equations can be solved exactly in such a field \cite{NN70,Grib91}
and the distribution function of created quasiparticles $f(\mathbf{p},t)$
can be found at any time as the solution of the following kinetic
equation \cite{1998e}
\begin{equation}\label{ke}
\frac{d f(\mathbf{p},t)}{d t} = \frac12
\Delta(\mathbf{p},t)\int\limits_{t_0}^t \! dt' \,
\Delta(\mathbf{p},t')\left[ 1 - 2f (\mathbf{p},t')\right]
\cos{\theta(\mathbf{p},t',t)},
\end{equation}
\begin{eqnarray}
\Delta(\mathbf{p},t)&= & e
E(t)\frac{\sqrt{m^2+p_{{\mbox{\tiny$\bot$}}}^2}}
{\varepsilon^2(\mathbf{p},t)},  \label{deltae}\\
\varepsilon(\mathbf{p},t)&=&\sqrt{m^2+p_{{\mbox{\tiny$\bot$}}}^2+
[p_{{\mbox{\tiny$\|$}}}-e A(t)]^2},\\
\theta(\mathbf{p},t',t)&=&2\int\limits_{t'}^{t}dt_1\,
\varepsilon(\mathbf{p},t_1),
\end{eqnarray}
where $e$ and $m$ are the particle charge and mass, $A(t)$ is the
vector potential of the electric field. The total field $E(t)$ is
defined as the sum of the external (laser) field $E_{ex}$ and the
self-consistent internal field $E_{in}$, which is determined by
Maxwell's equation
\begin{multline}\label{max}
\frac{d E_{in}(t)}{dt} = - \frac{e}{(2\pi)^3}\int \frac{d
\mathbf{p}}{\varepsilon(\mathbf{p},t)}\biggl\{
2\,p_{{\scriptscriptstyle\parallel}}\, f(\mathbf{p},t) \\
+ \sqrt{m^2+p_{\mbox{\tiny$\bot$}}^2} \int\limits_{t_0}^t \! dt_1
\, \Delta(\mathbf{p},t_1,t)\left[ 1 - 2 f(\mathbf{p},t_1)\right]
\times\cos{\left(
2\int\limits_{t_1}^{t}dt_2\,\varepsilon(\mathbf{p},t_2,t)\right)}
\biggr\}\ .
\end{multline}
The total current density on the r.h.s. of Eq.~(\ref{max}) is the
sum of the conductivity and vacuum polarization contributions,
respectively.

The equation \re{ke} was investigated carefully with reference to
the conditions in heavy ion collisions \cite{2001d} and in
laser fields \cite{2006a,Alkofer01,Roberts02}.
As opposed to the first case, where the characteristic times for
pair production $\tau_p \sim 1/m$ and the field action $\tau_f$
are of the same order, the laser field case corresponds to a
situation when these time scales differ by many orders of magnitude.
For example, $\tau_f/\tau_p \sim 10^{8}$ for optical lasers and
$\sim 10^4$ for X-ray free electron lasers \cite{Ringw01}.
As a consequence, the numerical task becomes extremely robust.

For the numerical investigation, equation \re{ke} is transformed
to a set of linear differential equations \cite{1998e}
\begin{eqnarray}\label{ode}
  \frac{df}{dt} = \frac{1}{2}\, \Delta \, u, \qquad
  \frac{du}{dt} = \Delta (1 - 2f_a)- 2 \varepsilon\, v, \qquad
  \frac{dv}{dt} = 2 \varepsilon\, u,
\end{eqnarray}
where $u, v$ are real auxiliary functions.
In such terms, the Maxwell equation \re{max} becomes
\begin{equation}\label{max1}
   \frac{d E_{in}(t)}{dt} = -  \frac{e}{(2\pi)^3}\int
\frac{d \mathbf{p}}{\varepsilon(\mathbf{p},t)}\biggl\{
2\,p_{{\scriptscriptstyle\parallel}}\, f(\mathbf{p},t) +
\sqrt{m^2+p_{\mbox{\tiny$\bot$}}^2}\ u \biggr\} .
\end{equation}
The system \re{ode} is integrated by the Runge-Kutta method with
the initial conditions $f(\mathbf{p},t_0)= u(\mathbf{p},t_0) =
v(\mathbf{p},t_0)=0$.
The envelope function in \re{s1} is chosen as
$\phi(t)=\theta(t)-\theta(t-\tau)$, where $\tau$ is the laser pulse duration.
The solution of \re{ke} is used to calculate the dynamical number density
\begin{equation}\label{dens}
n(t)= 2 \int\frac{d \mathbf{p}}{(2\pi)^3} f(\mathbf{p},t)
\end{equation}
and the annihilation rate in unit volume
\begin{equation}\label{num}
\frac{dN_a}{dV dt} = \int \frac{d\mathbf{p}_1}{(2\pi)^3}
\frac{d\mathbf{p}_2}{(2\pi)^3} \,
\sigma(\mathbf{p}_1,\mathbf{p}_2)f_a(\mathbf{p}_1,t)f_a(\mathbf{p}_2,t)
\times \sqrt{(\mathbf{v}_1 -\mathbf{v}_2)^2 - |\mathbf{v}_1 \times
\mathbf{v}_2 |^2} ,
\end{equation}
where $\mathbf{v}$ is the particle velocity and $\sigma$ the annihilation
cross-section.
The corresponding expression for spinor electrodynamics is \cite{Landau4}
\begin{equation}\label{spinor}
\sigma_e(\mathbf{p}_1,\mathbf{p}_2) = \frac{\pi e^4}{2m^2 \tau^2
(\tau-1)} \biggl[ \bigl(\tau^2 + \tau - 1/2 \bigr)
 \times \ln{\left\{ \frac{\sqrt{\tau} +
\sqrt{\tau-1}}{\sqrt{\tau} - \sqrt{\tau-1}}\right\} - (\tau+1)
\sqrt{\tau (\tau-1)} } \biggr].
\end{equation}
The t-channel kinematic invariant $\tau$ is given by
\begin{equation}\label{tau}
\tau = \frac{(p_1 + p_2)^2}{4m^2} = \frac{1}{4m^2}\bigl[
(\varepsilon_1 + \varepsilon_2)^2 - (\mathbf{p}_1 +
\mathbf{p}_2)^2 \bigr].
\end{equation}

\begin{figure}[t]
\centering \begin{minipage}{0.47\textwidth}
\includegraphics[height=70mm,keepaspectratio=true]{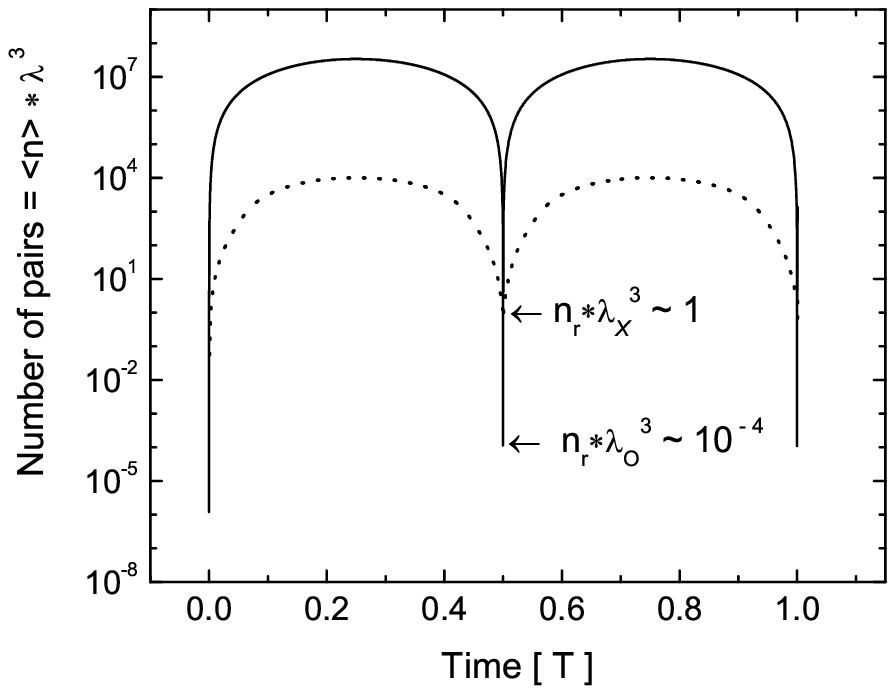}
\vspace*{-1mm}\caption{Time dependence of the quasiparticle number
in the volume $\lambda^3$ in the weak periodic field with the
parameters $E/E_{cr} = 2\cdot 10^{-5}$ and $\lambda = 800$ nm,
corresponding to the Jena Ti:sapphire laser \cite{Liesfeld05}
(solid line) and in the near-critical field case of an X-ray laser
\cite{Ringw01} with $E/E_{cr}=0.24$ and $\lambda =0.15$ nm
(dashed line).\label{fig3}}
\end{minipage} \hfill
\begin{minipage}{0.47\textwidth}\vspace*{7mm}
\includegraphics[width=0.95\textwidth,keepaspectratio=true]{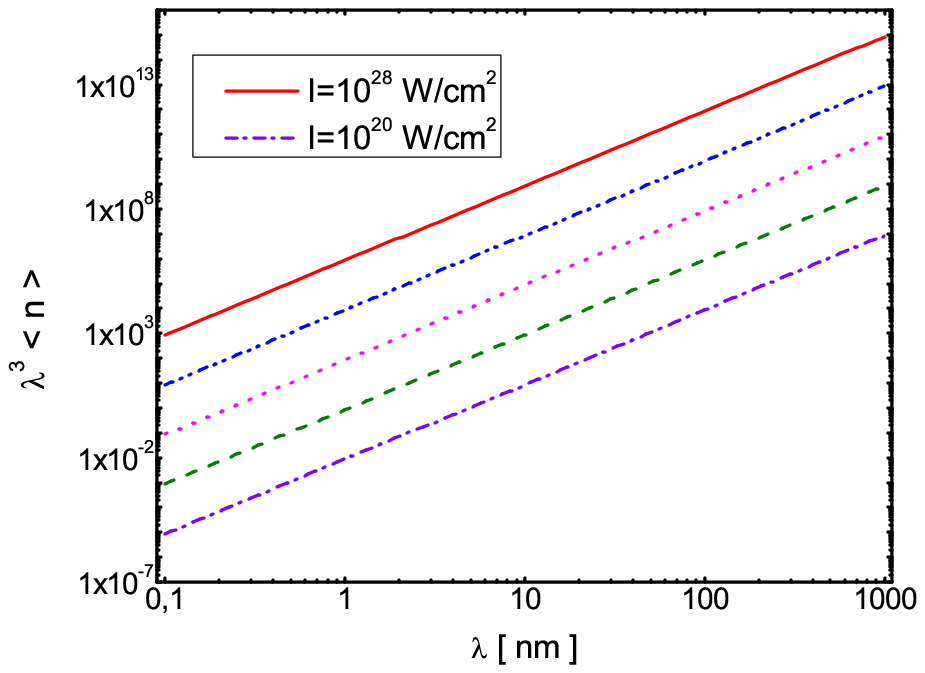}
\vspace*{5mm}\caption{ $\lambda$-dependence of the mean
quasiparticle number in the volume $\lambda^3$ for different field
strengths. The lines are drawn with a step width of $10^2$ W/cm$^2$.
The mean number density of pairs $<n>$ is independent of the wave
length in this range of parameters. \label{fig4}}
\end{minipage}
\end{figure}

Most of the previous works dealing with the estimation of pair
production in a laser field (e.g. \cite{Bunkin,Popov02}) have
considered the particles which remain after switching off the
external field (residual or true particles).
In this setting, the obtained results were negative:
the residual pair density is not observable for field strengths
$E \ll E_{cr}= m^2/e$.
The other possibility was investigated in Ref. \cite{2006a}.
The dynamical pair density during the action of laser pulse changes
periodically with twice the field frequency and its mean value
exceeds the residual density by many orders of magnitude, see Fig.
\ref{fig3}.
We suppose that these "quasi-particles" can undergo the standard processes
as, e.g., two-photon annihilation providing an indirect channel for
the observation of the dynamical Schwinger effect.

\section{Observable effects}

The peculiarity of the considered phenomenon is that the created pairs
are existing during the laser pulse only and decay almost
completely after the disappearance of the beams.
Nevertheless, in comparison with XFELs, modern optical lasers can generate
more pairs owing to the larger spot volume, and hence may provide now access
to observable signals of vacuum decay in indirect ways, such as the production
of coincident photon pairs from $e^+e^-$ annihilation (Fig. \ref{fig1})
or the change of the refraction index (Fig. \ref{fig2}).

The estimations made in Ref. \cite{2006a} in the framework of the
S-matrix formalism show that the operating lasers can produce in one
shot a few tens of $\gamma$-quanta, which can be detected by a
coincidence counter.
The rate of two-photon annihilation grows as $E^4$, therefore the planned
increase of laser intensity to $10^{24}-10^{25} $W/cm$^3$ will result in a
huge increase of the $2\gamma$ flux.
A further increase of the intensity to $10^{26}-10^{28} $W/cm$^3$ opens
the channels for creation of heavier particles such as charged muon and pion
pairs \cite{2006b} which annihilate into still harder $\gamma$-quanta with
energies in the range of $100-200$ MeV.

\begin{figure}[t]
\centering \centering
\begin{minipage}{0.47\textwidth}{
\includegraphics[width=0.99\textwidth,height=65mm]{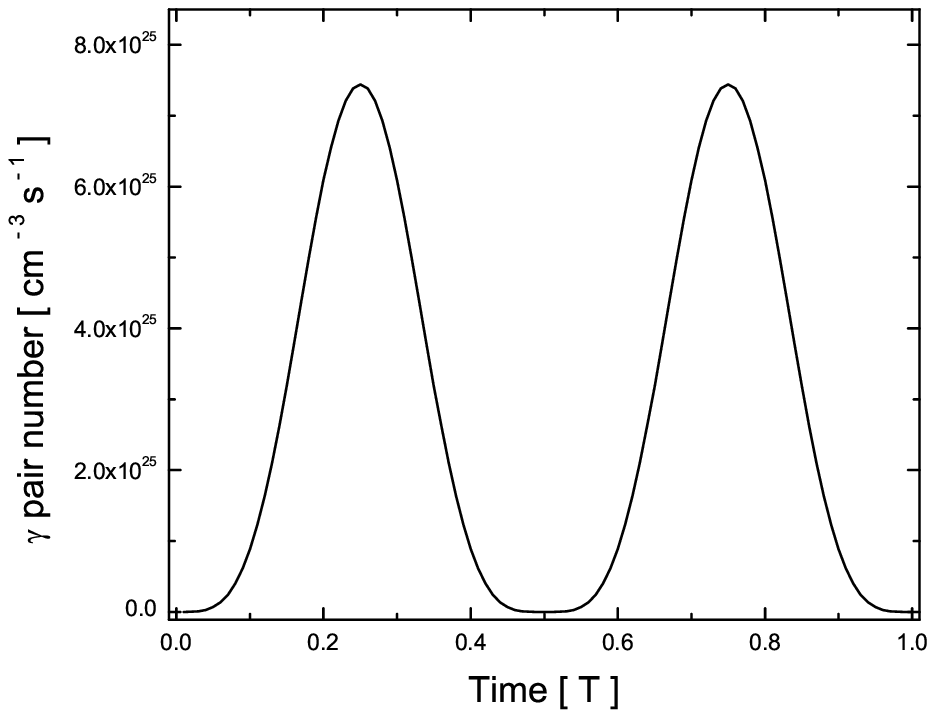}\hfill
\caption{The time dependence of annihilation rate (\ref{spinor})
of $e^+-e^-$ pairs in $\lambda_0^3$ volume for parameters of Jena
laser \cite{Liesfeld05}: I=10$^{20}$ W/cm$^2$, $\lambda_0=$ 800 nm.
\label{fig_ee}}}
\end{minipage}\hfill
\begin{minipage}{0.47\textwidth}{\vspace*{-1mm}
\includegraphics[width=0.9\textwidth,height=65mm]{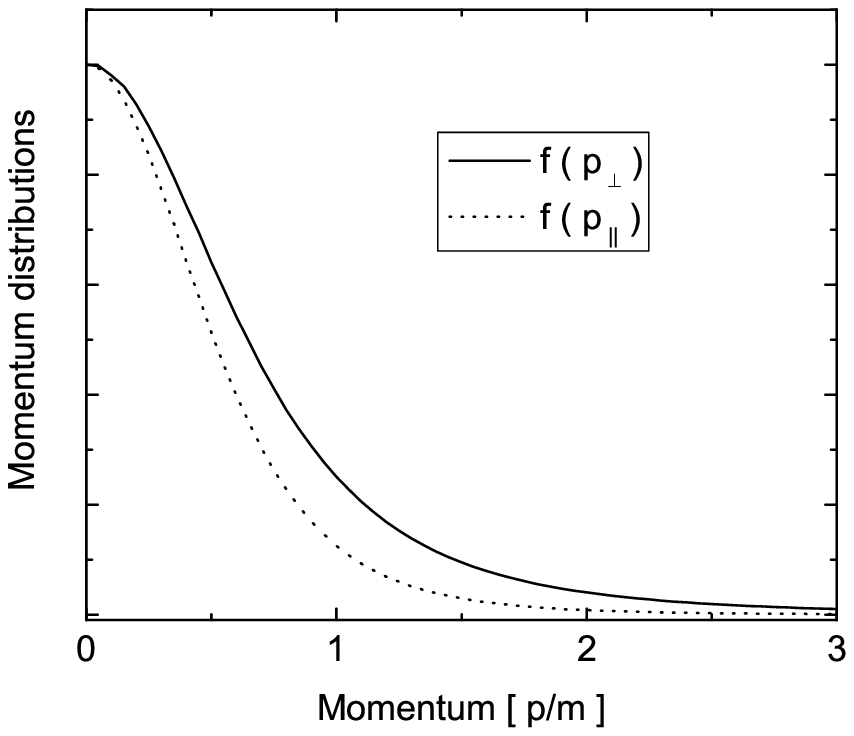}
\caption{The momentum dependence of created $e^+-e^-$ pairs.
\label{fig_pp}}}
\end{minipage}
\end{figure}

Another experimental method to reveal the $e^+e^-$-plasma  is to use
an additional probe laser for measuring the plasma refraction index,
see Fig. \ref{fig2}.
The collision of two counter-propagating laser beams with wavelength
$\lambda_0$ forms in the focus a standing wave with thin regions
("pancakes") filled by an electron-positron plasma with pulsing density
$n(t)\sim \sin^2{(\omega_0 t)}$.
The direction of the probe laser beam can be chosen in the range
$0-90^\circ$ which influences insignificantly on the effective path length
of beam in the presence of the plasma under typical conditions of high-power
lasers where the waist of focus is a few wave length and the pulse
duration is $10-100$ periods.
We consider the beam of a probe laser with short wavelength $\lambda_p \ll
\lambda_0$ propagating perpendicular to the direction of the
high-power laser beams.

According Fig. \ref{fig_pp}, the momentum spectrum of created
pairs is well modeled by a thermal equilibrium distribution with $T \sim m$ in the momentum
range $p \lsim m$.
Assuming also that the probe pulse duration is
short enough $\tau_p \ll \tau_0$, we can estimate the refraction
index in the quasi-stationary and high frequency approximations, respectively.
The problem reduces to a well-known one: the propagation of
transverse waves in an equilibrium plasma with a Debye length $r_D
\sim 10^{-5} cm < \lambda_0$ and a Langmuir frequency $\omega_L
\sim 10^{14}/c$.

In the high frequency approximation $\omega_p \gg kv_T$, the real part of the
transverse dielectric permeability \cite{Landau10} is
\begin{equation}\label{2}
    \varepsilon_{tr}= 1- \frac{\omega^2_L}{\omega_p^2}\left[
    1+\left(\frac{kv_T}{\omega_p}\right)^2\right].
\end{equation}
The solution of the dispersion equation of transverse waves can be
written as
    \begin{equation}\label{3}
\omega_p^2 = \omega^2_L +k^2 \left[1+ \frac{\omega^2_L}{\omega^2_L +
k^2}\right]~,
\end{equation}
and the refraction index is
    \begin{equation}\label{4}
    n= \sqrt{\frac{1-\eta^2}{1+\eta^2}}~,
\end{equation}
where $\eta = \omega_L/\omega_p$.
It follows  that $n=0.999975$ for $\omega_p=10\,\omega_0$  and
$n=0.99999975$ for $\omega_p = 100\,\omega_0$.
Such small differences of refraction indices of the split probe beams
appears to be a challenge for the experimental detection by an interference pattern
because of the rather short path in the plasma.
To comfortably detectable, the value of the refraction index of the plasma should increase
by about two orders of magnitude so that a clear interference pattern emerges.
The increase of the laser intensity to I=10$^{22}$-10$^{23}$ W/cm$^2$ will
apparently be sufficient to create the conditions for observability of an
interference pattern for the high-frequency probe beams.
{The more comprehensive calculation of the refraction index is performed in the work \cite{Blaschke:2008wf}.}

\begin{figure}[t]
\centering
\includegraphics[width=0.47\textwidth,keepaspectratio=true]{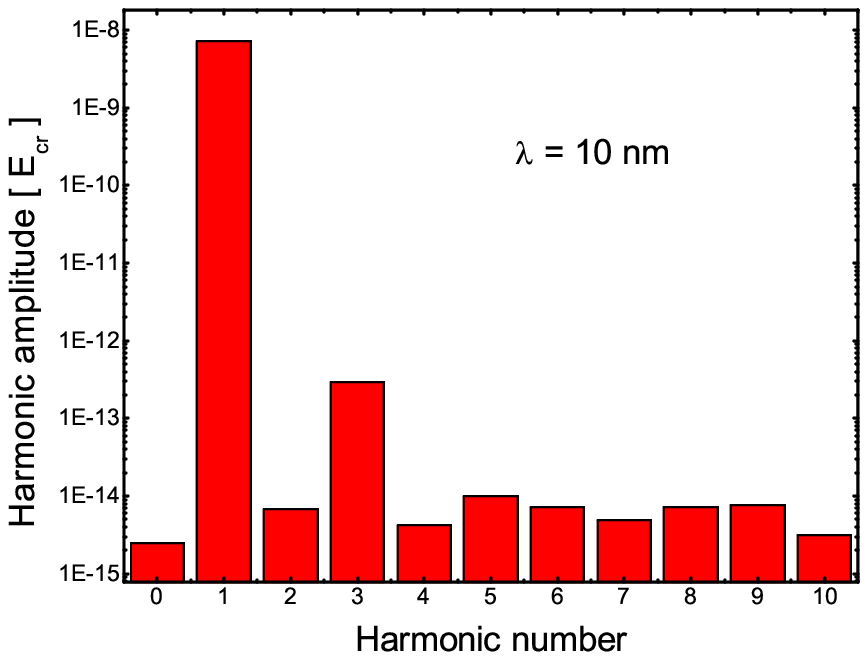}\hfill
\includegraphics[width=0.47\textwidth,keepaspectratio=true]{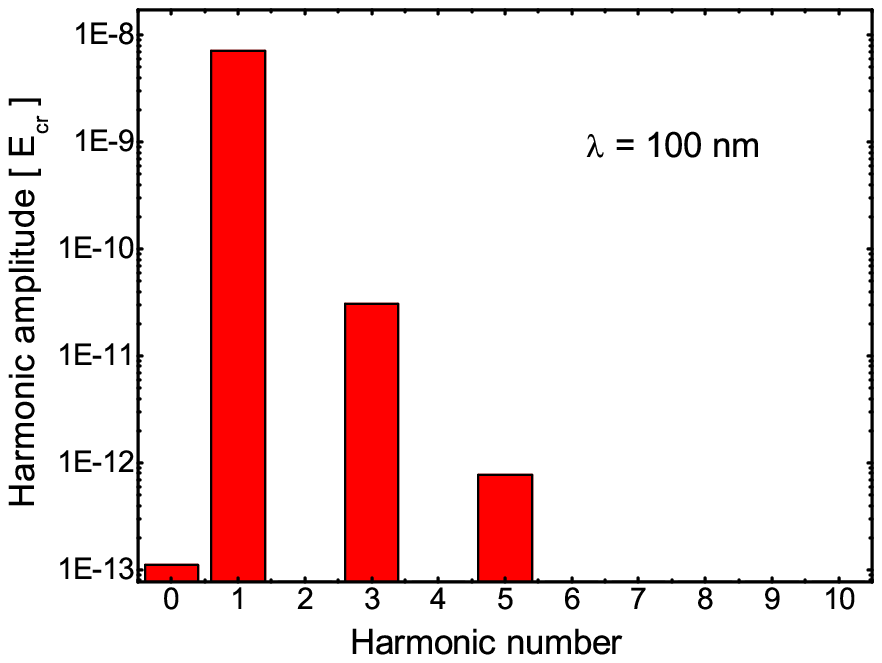}
\caption{The dependence of the amplitudes of higher harmonics in the
back-reaction field on the harmonics number. \label{fig_hh}}
\end{figure}

The registration of higher harmonics is one more perspective method for the
experimental observation of non-linear QED effects in a strong laser field.
Here we investigate the spectrum of oscillations of the back-reaction field
(\ref{max}).
Fig.~\ref{fig_hh} shows that the third harmonics of the laser frequency
$\omega_0$ is the most intensive one and its amplitude increases together
with the laser wave length.
The expected value of the amplitude for the third harmonics reaches the order
of 10$^6$ V/cm for a laser intensity of 10$^{20}$ W/cm$^2$.

\section{Conclusion}

Our estimates show that in the field of optical lasers with an
intensity of $10^{28}$~W/cm$^2$ one can expect $\sim 10^{16}$
$e^+-e^-$ annihilation events per laser pulse accompanied by irradiation
of $\sim 1$ MeV $\gamma$-quanta which can be registered outside the focus
of the counter-propagating laser beams by coincidence counters.
The expected amplitude of the third harmonics of the laser frequency reaches
the order of 10$^6$ V/cm at a laser intensity of 10$^{20}$ W/cm$^2$.
The interference scheme for the observation of a refraction index of
the transient $e^+-e^-$ plasma can become feasible by a further increase of
the laser intensity up to 10$^{23}$-10$^{24}$ W/cm$^2$.

\end{document}